\documentclass[preprint,review,12pt]{elsarticle} 
\usepackage{amssymb, amsmath, psfrag, ae, graphicx, placeins,longtable, cleveref}
\usepackage[hang]{subfigure}
\usepackage[mathscr]{euscript}
\usepackage{units,eurosym,booktabs, color, booktabs, subfloat, xfrac}
\usepackage{changebar}
\def\ve#1{\mathchoice{\mbox{\boldmath$\displaystyle#1$}}
{\mbox{\boldmath$\textstyle#1$}}
{\mbox{\boldmath$\scriptstyle#1$}}
{\mbox{\boldmath$\scriptscriptstyle#1$}}} 
\newcommand{\dd}{\text{d}}
\newcommand{\dt}{\dd t}
\newcommand{\ff}{\text f}
\newcommand{\wrt}{\text{w.r.t. }}

\newcommand{\st}{\text{s.t.}}

\newcommand{\eg}{\text{e.g., }}
\newcommand{\ie}{\text{i.e., }}

\journal{Computers \& Chemical Engineering}

\begin{document}
\begin{frontmatter}
\title{Dynamic Real-time Optimization of Batch Processes using Pontryagin's Minimum Principle and Set-membership Adaptation}
\author[FCFT, cor]{Radoslav Paulen}
\ead{radoslav.paulen@stuba.sk}
\author[FCFT]{Miroslav Fikar}
\cortext[cor]{Corresponding author. Tel.: +421 259 325 730; fax: +421 259 325 340.}
\address[FCFT]{Faculty of Chemical and Food Technology, Slovak University of Technology in Bratislava, Radlinskeho 9, Bratislava, Slovakia}

\begin{abstract}
This paper studies a dynamic real-time optimization in the context of model-based time-optimal operation of batch processes under parametric model mismatch. In order to tackle the model-mismatch issue, a receding-horizon policy is usually followed with frequent re-optimization. The main problem addressed in this study is the high computational burden that is usually required by such schemes. We propose an approach that uses parameterized conditions of optimality in the adaptive predictive-control fashion. The uncertainty in the model predictions is treated explicitly using reachable sets that are projected into the optimality conditions. Adaptation of model parameters is performed online using set-membership estimation. A class of batch membrane separation processes is in the scope of the presented applications, where the benefits of the presented approach are outlined.
\end{abstract}

\end{frontmatter}

\section{Introduction}
\label{sec:introduction}
Optimization of operations of batch processes is a rich field of research. One of the main goals is to reduce the variability among the produced batches despite the present uncertainties and disturbances. This problem struck the attention of many research groups~\citep{nag03, sri03b, ade09, fra13, luc13, mar15, jan16}.

In this paper, we consider a real-time implementation of a control policy under parametric plant-model mismatch that optimizes a batch process by assigning dynamic degrees of freedom such that a certain performance index is optimized. Similar problems were studied in many previous works using on-line or batch-to-batch adaptation of the optimality conditions~\citep{fra05, fra13}, by mid-course correction~\citep{yab97, hos13} or by design of robust controller for tracking the conditions of optimality~\citep{nag03}. Another set of approaches to the problem uses advanced robust strategies in the framework of model predictive control~\citep{luc13}. This paper proposes an adaptation of these approaches to the problem of dynamic real-time optimization of batch processes. This task is not straightforward because if one uses a receding-horizon control strategy, the prediction horizons used need to be quite long, because of the usual presence of terminal constraints, which might compromise the real-time feasibility of the scheme.

We base the presented methodology on the parameterization of the optimal operation using the optimality conditions given by Pontryagin's minimum principle. As the cost is usually insensitive \wrt a precise singular control trajectory~\citep{sri03b}, the parameterization of the optimal policy makes the real-time decision problem to mainly boil down to identification of switching times of the optimal control policy. Such approach reduces computational burden while allowing for the use of sufficiently long prediction horizons when projecting the parametric uncertainty in controller performance and feasibility, particularly \wrt terminal time conditions. Robustness \wrt parametric uncertainty is addressed by taking into account the imprecision of parameter estimates, which is projected into the uncertainty of the switching times. In order to improve performance of such a controller, \ie to reduce conservatism introduced by uncertain switching times, we use on-line parameter estimation. While having the optimal control policy explicitly parameterized in the uncertain parameters, one can tailor the real-time implementation of the optimal operation, e.g., in a way that minimizes the number of on-line calculations.

The novelty of this paper lies foremost in an effective combination of Pontryagin's minimum principle and set-based techniques (set-membership estimation and reachability analysis). This gives rise to a methodology capable of projecting the propagation of the uncertainty in model parameters into uncertainty in the optimal operation of a plant. Using this methodology, efficient and effective real-time optimization of a plant can be established.

The outline of the paper is as follows. Section~\ref{sec:prelim} present preliminary theoretical knowledge on Pontryagin's minimum principle~\citep{pon62} and on set-membership estimation~\citep{schw68, fog82}. The former is used to parameterize the optimality conditions of the dynamic optimization problem, while the latter technique is used for adaptation of the model parameters based on the measured data along the process run. Next we propose the implementation of the real-time optimization using parameter adaptation. Finally, we present a case study from chemical engineering domain and discuss various aspects of the obtained results.

\section{Problem definition}
In this paper, we consider a real-time implementation of a control policy that optimizes a process by assigning dynamic degrees of freedom such that a certain performance index is optimized:
\begin{subequations}\label{eq:prob_gen}
\begin{align}
\min_{u(t), t_\ff} & \ \mathcal J(\ve p) \!:= 
\min_{u(t), t_\ff} \int_{0}^{t_\ff} \!F_0(\ve x(t, \ve p), \ve p) + F_u(\ve x(t, \ve p), \ve p) u(t)\,\dt\\
\st \ & \dot{\ve x}(t, \ve p) = \ve f_0(\ve x(t, \ve p), \ve p)
+ \ve f_u(\ve x(t, \ve p), \ve p) u(t),\label{eq:model}\\
& \ve x(0) = \ve x_0, \quad \ve x(t_\ff, \ve p) = \ve x_\ff, \\
& u(t)\in [u^L, u^U],
\end{align}
\end{subequations}
where $t$ is time with $t\in[0, t_\ff]$, $\ve x(\cdot)$ is an $n$-dimensional vector of state variables, $\ve p$ is an $m$-dimensional vector of model parameters, $u(t)$ is a (scalar) manipulated variable, $F_0(\cdot)$, $F_u(\cdot)$, $\ve f_0(\cdot)$, and $\ve f_u(\cdot)$ are continuously differentiable functions, $\ve x_0$ represents a vector of initial conditions, and $\ve x_\ff$ are specified final conditions. We note here that an inclusion of multi-input and/or state-constrained cases is a straightforward extension but it is not considered in this study for the sake of simplicity of the presentation. We also note that the specific class of input-affine systems is a suitable representation for a large variety of the controlled systems~\citep{han06}. For a general nonlinear model, one may use simple manipulations to rearrange the model into input-affine structure~\citep{son98}, which might though increase the number states of the problem. In the domain of chemical engineering, it is, however, very common to encounter input-affine problems~\citep{amr10} (e.g., when the optimized variable is a reactor feed) or to reformulate the model and arrive at the input-affine structure~\citep{lio95}.

We will assume that the plant behavior is known qualitatively and that the only source of uncertainty is present in the unknown values of model parameters. Only a prior knowledge is assumed about the parameters, i.e., the true values of the parameters lie in the a priori known interval box $\ve P_0:=[\ve p_0^L, \ve p_0^U]$, where superscripts $L$ and $U$ denote the lower and upper bounds of $\ve p$. The nominal realization of the uncertainty will be assumed as $\ve p^\text{nom}:=\text{mid}(\ve P)$, where $\text{mid}(\cdot)$ indicates a mid-point of the interval box.

We will also assume that certain measurements are available from the plant. Their corresponding model-based predictions are
\begin{align}\label{eq:output}
 \ve y(t) = \ve g(\ve x(t,\ve p), \ve p),
\end{align}
where $\ve g(\cdot)$ is a continuously differentiable vector function.

\section{Preliminaries}\label{sec:prelim}
\subsection{Conditions for Optimality}\label{ssec:pmp}
Pontryagin's minimum principle can be used~\citep{joh63, sri03a, pau12jms, pau15jpc} to identify the optimal solution to~\eqref{eq:prob_gen} via enforcing the necessary conditions for minimization of a Hamiltonian
\begin{align}\label{eq:hamiltonian}
 H:= \mu^L(u^L-u) + \mu^U(u-u^U) + \underbrace{F_0
 +\ve\lambda^T\ve f_0}_{H_0(\ve x(t, \ve p), \ve \lambda(t, \ve p), \ve p)} \!+ \underbrace{\left(F_u+\ve\lambda^T\ve f_u \right)}
 _{H_u(\ve x(t, \ve p), \ve \lambda(t, \ve p), \ve p)}u,
\end{align}
where $\ve\lambda(\cdot)$ is a vector of adjoint variables, which are defined through
\begin{align}
&\dot{\ve\lambda}(t, \ve p)=-\frac{\partial H}{\partial \ve x}(t, \ve p),
\quad \ve\lambda(t_\ff, \ve p)=\ve\nu(\ve p),
\end{align}
and $\mu^L(t,\ve p)$, $\mu^U(t,\ve p)$, and $\ve\nu(\ve p)$ are the corresponding Lagrange multipliers.
The optimality conditions of~\eqref{eq:prob_gen} can then be stated as~\citep{sri03a}: $\forall t\in[0, t_\ff]$,
\begin{align}
 \frac{\partial H}{\partial u}:=
 H_u(\ve x(t,\ve p), \ve \lambda(t,\ve p),  \ve p) - \mu^L(t,\ve p) + \mu^U(t,\ve p) &= 0,\label{eq:optcon_u}\\
 H(\ve x(t,\ve p), \ve \lambda(t,\ve p), \ve p, u(t), \mu^L(t,\ve p), \mu^U(t,\ve p))&=0, \\
 H_0(\ve x(t,\ve p), \ve \lambda(t,\ve p), \ve p)=0, \quad \ve x(t_\ff,\ve p) - \ve x_\ff &=0.
\end{align}
The condition $H=0$ arises from the transversality, since the final time is free~\citep{pon62}, and from the fact that the optimal Hamiltonian
is constant over the whole time horizon, as it is not an explicit function of
time. The condition $H_0=0$ is the consequence of the former two conditions. Since the Hamiltonian is affine in input (see~\eqref{eq:hamiltonian}), the optimal trajectory of control variable is either determined by active input constraints or it evolves inside the feasible region.

Assume that for some point $t$ we have $H_u=0$ and $u^L<u(t)<u^U$. It follows from~\eqref{eq:optcon_u} that the optimal control maintains $H_u(\cdot)=0$. Such control is traditionally denoted as singular. Further properties of the singular arc, such as switching conditions or state-feedback control trajectory can be obtained by differentiation of $H_u$ with respect to time (sufficiently many times) and by requiring the derivatives to be zero. The time derivatives of $H$ and $H_0$ must be equal to zero as well. Earlier results on derivation of optimal control for input-affine systems~\citep{sri03a} suggest that it is possible to eliminate $\ve\lambda(\cdot)$ from the optimality conditions and thus to arrive at analytical characterization of switching conditions between singular and saturated-control arcs.

As the optimality conditions obtained by the differentiation \wrt time are linear in the adjoint variables, the differentiation of $H_u$ (or $H_0$) can be carried out until it is possible to transform the obtained conditions to a pure state-dependent switching function $S(\ve x(t), \ve p)$. It is usually convenient to use a determinant of the coefficient matrix of the equation system $\ve A\ve\lambda=\ve0$ for this. The singular control $u_\text s(\ve x(t), \ve p)$ can be found from differentiation of switching function \wrt time as
\begin{align}\label{eq:singular_ctrl_gen}
 \frac{\dd S}{\dt} =
 \frac{\partial S}{\partial \ve x^T}\frac{\dd \ve x}{\dt}=&
 \ \frac{\partial S}{\partial \ve x^T}(\ve f_0 +\ve f_u u_\text s) =0\notag\\
 \quad \Rightarrow \quad
 u_\text s (\ve x(t,\ve p), \ve p)=& -\frac{\partial S}{\partial \ve x^T}
 \ve f_0\bigg/
 \frac{\partial S}{\partial \ve x^T}\ve f_u.
\end{align}

The resulting optimal-control policy is then given as a step-wise strategy~\citep{pau15jpc} by
\begin{align}\label{eq:singular_ctrl_switch}
u^\ast(t, \ve\pi) :=
\begin{cases}
  u^L, &  t\in[0, t_1), \ S(\ve x(t,\ve p), \ve p) >0,\\
  u^U, & t\in[0, t_1), \ S(\ve x(t,\ve p), \ve p) <0,\\
  u_\text s(\ve x(t,\ve p), \ve p), & t\in[t_1, t_2), \ S(\ve x(t,\ve p), \ve p) =0,\\
  u^L, & t\in[t_2,t_\ff], \ S(\ve x_\ff, \ve p) <0,\\
  u^U, & t\in[t_2,t_\ff], \ S(\ve x_\ff, \ve p) >0,
\end{cases}\\
\ve x_\ff=\ve x(t_2, \ve p) + \int_{t_2}^{t_\ff}\!\ve f_0(\ve x(t,\ve p), \ve p) + \ve f_u(\ve x(t,\ve p), \ve p) u^\ast(t, \ve\pi)\,\dt,\label{eq:fin_cond}
\end{align}
where $\ve\pi:=(\ve p^T, t_1, t_2, t_\ff)^T$ is the vector that parameterizes the optimal control strategy. Note that the presented optimal-control strategy determines implicitly the switching times $t_1$, $t_2$ and the terminal time $t_\ff$ as functions of model parameters $\ve p$.

In case that the use of the minimum principle turns out to be too complex (e.g., many derivatives are needed to characterize the solution), a numerical identification of the control arcs~\citep{sch05, sch06} or
a recently presented parsimonious input parameterization~\citep{erd18, rod19} can be used.

\subsection{Set-membership estimation}\label{ssec:sme}
In order to estimate the model parameters, we will make use of plant outputs (measurements), whose predictions are expressed as in~\eqref{eq:output}. We will assume that the true output of the plant $\ve y_p(t)$ is corrupted with a (sensor) noise that is bounded with a known magnitude $\ve\sigma$. Thus, the measured output $\ve y_m(t)$ is such that
\begin{align}
 |\ve y_m(t) - \ve y_p(t)|\leq\ve\sigma,
\end{align}
where the absolute value is understood component-wise. In turn, the
set-membership constraints for predicted output $\ve y(t)$ apply in the form:
\begin{align}\label{eq:bound_meas2}
 |\ve y_m(t) - \ve y(t)|\leq\ve\sigma.
\end{align}

We are interested in the determination of parametric bounds such that
\begin{equation}
 \ve P_k \subseteq \ve P_{k-1} \subseteq \dots \subseteq \ve P_1 \subseteq \ve P_0,
\end{equation}
where $k$ is the ordinal number of a measurement taken. The parametric bounds can be determined through solution of a series of optimization problems as~\citep{mupa_cace17, wal18}:
\begin{subequations}\label{eq:nlp_par_bounds}
\begin{align}
p_{k,j}^L/&p_{k,j}^U := \min_{\ve p\in\ve P_0} \ /\max_{\ve p\in\ve P_0} p_j\\
\text{s.t. }& \dot{\ve x}(t, \ve p) = \ve f_0(\ve x(t, \ve p), \ve p)
+ \ve f_u(\ve x(t, \ve p), \ve p) u(t),\ \forall t\in[0,t_k],\\
&\ve x(0, \ve p) = \ve h (\ve p),\\
&\ve y(t_i, \ve p) = \ve g (\ve x(t_i, \ve p), \ve p),\ \forall i \in \{1, \dots, k\},\\
&-\ve\sigma \leq \ve y(t_i, \ve p) - \ve y_m(t_i) \leq \ve\sigma,\ \forall i \in \{1, \dots, k\}, \label{eq:nlp_par_bounds_err}
\end{align}
\end{subequations}
for given $u(t)$, where $j\in\{1,\dots, n_p\}$ indicates the $j^\text{th}$ element of a vector.

\section{Dynamic real-time optimization}
As the optimal control structure is a function of uncertain parameters, the uncertainty should be taken into account when devising a real-time implementation of the optimal control of the plant.

\subsection{Projection of parametric uncertainty into solution strategy}\label{ssec:param}
\begin{figure}
\centering
\psfrag{S}{\hspace{-0.6cm} $S(\cdot)$}
\psfrag{time}{$t$}
\psfrag{0}{\hspace{-0.2cm} 0}
\psfrag{1}[cc][bb]{\hspace{-0.5cm} 0}
\psfrag{t1}[t][b][1][0]{\hspace{-0.1cm}$[t^L_1(\ve P), t^U_1(\ve P)]$}
\includegraphics[width=0.8\linewidth]{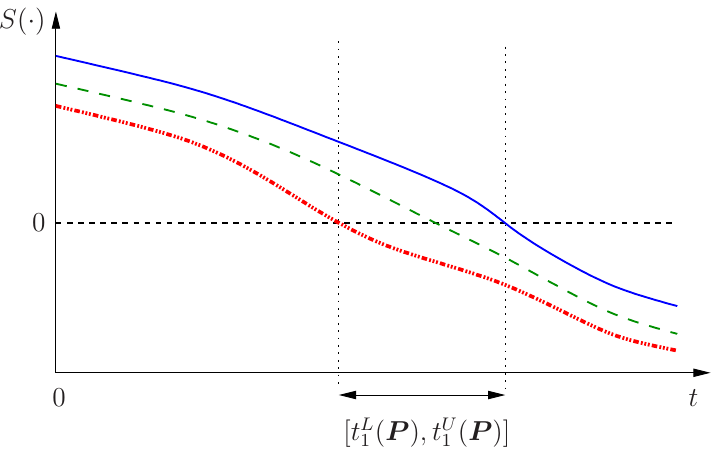}
\caption{Illustration of the switching function evolution under uncertainty using (set-based) reachability analysis with nominal uncertainty realization (green dashed line) and extreme realizations (dash-dotted red and solid blue lines).}
\label{fig:sparam}
\end{figure}

Given the structure of the optimal-control policy~\eqref{eq:singular_ctrl_switch}, one can project the parametric uncertainty into uncertainty of the switching times and singular control as ($\forall \ve p \in \ve P$)
\begin{subequations}\label{eq:bound_opt}
\begin{align}
t_l(\ve p)&\in[t_l^L(\ve P), t_l^U(\ve P)]=:T_l, \ \forall l\in\{1, 2, \ff\},\\
u_s(t, \ve p)&\in[u_s^L(t, \ve P), u_s^U(t, \ve P)]=:U^\text{opt}(t), \forall t\in[t_1(\ve p), t_2(\ve p)].
\end{align}
\end{subequations}
This can be achieved either by using some set-theoretic techniques for calculating reachable sets~\citep{cha15} or by sampling approaches. Figure~\ref{fig:sparam} provides an illustration, where the reachable sets are shown over time for the switching function $S(\cdot)$.

Formally, the problem of determination of~\eqref{eq:bound_opt} can be cast as a set-inversion problem~\citep{jau93}. As an example, let us consider that $S(\ve x(0,\ve p), \ve p)>0, \ \forall \ve p\in\ve P_0$. The interval $T_1$ can then be defined as:
\begin{equation}
 T_1:=\left\{t_1 \,\left|\,
 \begin{array}{c}
  \exists \ve p\in \ve P, \forall t\in[0,t_1]:\\
  \ve x(0, \ve p) = \ve h (\ve p),\ S(\ve x(t_1,\ve p), \ve p) = 0,\\
  \dot{\ve x}(t, \ve p) = \ve f_0(\ve x(t, \ve p), \ve p)
+ \ve f_u(\ve x(t, \ve p), \ve p) u^L 
 \end{array}\right.
\right\}.
\end{equation}
The rest of the uncertain intervals and controls can be defined and determined analogously. Efficient set-inversion techniques exist~\citep{pau15-ima} and can be used herein. The problem might also be reformulated to a bound-determining optimization problem, similarly to the estimation problem in~\eqref{eq:nlp_par_bounds}. Using such a reformulation, it is also possible to merge the problems of reachability analysis and set-membership estimation and to formulate the reachability-analysis problem directly over the collected data. This way several possible deficiencies (such as those arising from outbounding the parameter set by a box) can be eliminated.

\begin{figure}
\centering
\psfrag{alpha}{\hspace{-0.4cm}$u^\text{opt}(t,\cdot)$}
\psfrag{time}{$t$}
\psfrag{0}{\hspace{-0.4cm} $u^L$}
\psfrag{1}{\hspace{-0.5cm} $u^U$}
\psfrag{t1}[t][b][1][0]{\hspace{-0.1cm}$[t^L_1(\ve P), t^U_1(\ve P)]$}
\psfrag{t2}[t][b][1][0]{\hspace{-0.4cm}$[t^L_2(\ve P), t^U_2(\ve P)]$}
\psfrag{t3}[t][b][1][0]{\hspace{0.4cm}$[t^L_\ff(\ve P), t^U_\ff(\ve P)]$}
\includegraphics[width=0.8\linewidth]{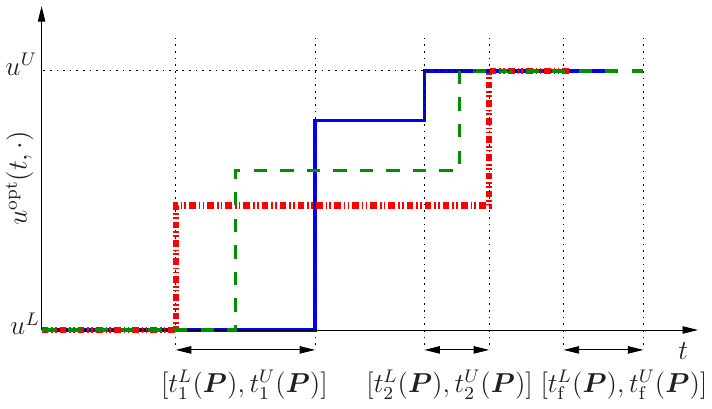}
\caption{Illustration of the parameterization of the optimal control policy under uncertainty with nominal policy (green dashed line) and extreme-case realizations (dash-dotted red and solid blue lines).}
\label{fig:uparam}
\end{figure}

Figure~\ref{fig:uparam} illustrates the parameterization~\eqref{eq:bound_opt} for a simple case, where the singular control is constant. Note that the parameterization reveals time intervals (i.e., $[0, t_1^L(\ve P)]$, $[t_1^U(\ve P), t_2^L(\ve P)]$, and $[t_2^U(\ve P), t_\ff^L(\ve P)]$), which are parts of the optimal solution for any realization of uncertain parameters and are thus invariant to the presence of uncertainty. 

A particular technical advantage can be exploited for determination of the switching intervals \ie that the integration in~\eqref{eq:fin_cond} can be done backwards in time from the final condition. As the batch processes exhibit inherently unstable dynamics, their backward integration is stable~\citep{cao03}. Such a feature can readily be exploited by modern reachability analysis approaches for parametric ordinary differential-algebraic equations~\citep{vil15}.

\subsection{Robust approach to real-time optimization}\label{sec:rob}
The result~\eqref{eq:bound_opt}, in practice, establishes a parametric solution to the real-time optimization problem. Its implementation can be performed in a robust fashion to determine the parameters of the optimal-control structure that lead to the best performance in the worst case. We can then solve
\begin{align}\label{eq:prob_rob_nco}
 \min_{\stackrel{u_s(t, \ve p)\in U^\text{opt}(t), \forall t\in[t_1(\ve p), t_2(\ve p)]}
 {t_l\in T_l, \ \forall l\in\{1, 2, \ff\}}}
 \max_{\ve p\in\ve P_0} \ \|\mathcal J(\ve p) - \mathcal J(\ve p^\text{nom}_0)\|_2^2
 \qquad \st \ \eqref{eq:model}, \eqref{eq:singular_ctrl_switch}, \eqref{eq:fin_cond},
\end{align}
for a given $\ve x(0) = \ve x_0$ and $\ve P_0$. Here we propose to minimize the variance of the objective \wrt nominal scenario under the worst-case realization of $\ve p \in\ve P$, which can also be modified to $\|\mathcal J(\ve p) - \min_{p^\text{opt}\in\ve P_0} \mathcal J(\ve p^\text{opt})\|_2^2$. Note that this goal goes in line with the efforts of practical batch process control, where the reduction of the batch-to-batch variability is one of the main targets of decision making.

\subsection{Robust adaptive approach to real-time optimization}
\label{sec:adapt}
In order to reduce conservatism of a robust scheme, parameter estimation can be used for exploitation of data gathered along the process run. The employed parameter estimation scheme should take into account the presence of noise in the measurements. Here we propose to use set-membership strategy outlined in Section~\ref{ssec:sme}.

The problem~\eqref{eq:prob_rob_nco} can then be resolved with the initial state conditions $\ve x(k) = \ve x_k$ and with updated parameter bounds $\ve P_k$ in a shrinking-horizon fashion. Computational efficiency of this real-time optimization scheme can be achieved by exploiting the fact that the re-optimization does not need to be done at each sampling time of the plant (i.e., when new measurements become available) but can be scheduled before a consecutive switching event must be realized. As an example, consider Fig.~\ref{fig:uparam}, where one can start the operation on the lower-bound of the input variable and estimation of parameter bounds and re-optimization can be scheduled in the sampling instant of the plant just before minimal value of the time $t_1$, $t_1^L(\ve P)$. The re-optimization with updated bounds on parameters would then update (possibly increase) the value of $t_1^L(\ve P)$. Further re-optimizations can then follow based on this scheme.

Once the optimal value of the objective function of~\eqref{eq:prob_rob_nco} reaches $\|\mathcal J(\ve P) - \mathcal J(\ve p_k^\text{nom})\|_2^2<\varepsilon$, where $\varepsilon>0$ represents user-defined tolerance for the worst-case cost variation, the calculated control actions can be implemented until the end of the batch, \eg with a feedback scheme~\citep{fra13}, until the terminal conditions are met.

Note that because of the switching nature of the optimal control strategy, the
proposed problem might show discontinuity when the set of active constraints
changes. This can be remedied by the adaptation of continuous-formulation
technique presented in~\cite{pra11}.

A pseudo-algorithm can be devised at this point to summarize the proposed approach:
\begin{enumerate}
 \item Given the problem setup (Eqs.~\eqref{eq:prob_gen} and~\eqref{eq:output}), identify the solution structure using Pontryagin's minimum principle (as shown in Section~\ref{ssec:pmp}).
 \item Given $\ve P_0$, use reachability analysis to project the uncertainty in the parameters to uncertainty about the solution structure (as shown in Section~\ref{ssec:param}).
 \item Apply the optimal policy until the next uncertain switching time and collect the measurements along. \label{step:apply}
 \item Solve problem~\eqref{eq:nlp_par_bounds} to determine new interval box $\ve P$ and re-calculate the uncertain solution structure.\label{step:estim}
 \item If significant reduction in the uncertainty of the switching time is achieved, go to Step~\ref{step:apply}. Else solve problem~\eqref{eq:prob_rob_nco} to determine the switching times. Apply the first switching in the control, collect the data along and go to Step~\ref{step:estim}.
\end{enumerate}

\section{Case study}
\label{sec:process_desc}
We demonstrate the findings of this study on an example of time-optimal control of a batch diafiltration process~\citep{che98}. This is a membrane-based separation process designed for a simultaneous concentration of valuable products in the liquid solutions (referred to as macro-solutes) and a wash-out of the impurities (referred to as micro-solute).

\begin{figure}
\centering
\includegraphics[width=0.6\linewidth]{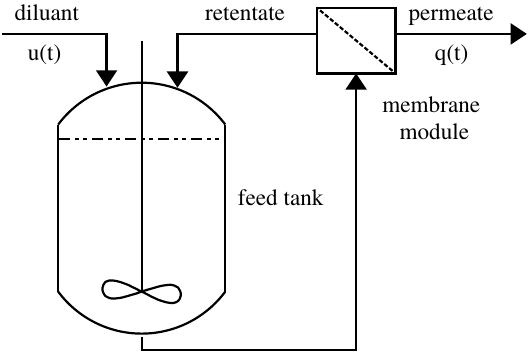}
\caption{Schematic representation of a generalized diafiltration process.}
\label{fig:plant_scheme}
\end{figure}

A simplified scheme of the plant is shown in Fig.~\ref{fig:plant_scheme}. 
After the separated solution with initial volume ($V_0$) comprising a macro-solute (high molecular weight component) and a micro-solute (low molecular weight component) with initial concentrations $c_{1,0}$ and $c_{2,0}$, respectively, is transferred to the feed tank, the operation of the process is launched. Solution containing diluant (solvent), micro-solute and macro-solute is taken from the feed tank to the membrane module. The installed membrane is designed in a way to allow passage of micro-solute and to retain macro-solute. Permeate stream then leaves the system with the flow rate $q$ which is specific for a given membrane, operating conditions and is often a function of actual concentrations of separated species. Retentate stream is then introduced back into the feed tank. Once the final conditions, which are the prescribed final concentrations of the species $c_{1,\ff}$ and $c_{2,\ff}$, are met, the process is terminated and the solution is withdrawn from the system.
During the operation, the transmembrane pressure is controlled at a constant value. The temperature of the solution is maintained around a constant value using a heat exchanger (not shown in Fig.~\ref{fig:plant_scheme} for the sake of simplicity). The manipulated variable $u(t)$ is the ratio between fresh water inflow into the tank and the permeate outflow $q$.

In our study, the outflow $q$ is measured at intervals of one minute and its model is given by
\begin{align}\label{eq:ext_lim_model}
q(\underbrace{c_1(t), c_2(t)}_{\ve c(t)}, \underbrace{\gamma_1, \gamma_2, \gamma_3}_{\ve\gamma}) &=  A\gamma_1 \ln\left(\frac{\gamma_2}{c_1(t) c_2^{\gamma_3}(t)}\right)\notag\\
    &=\notag
     A\gamma_1\left[\ln(\gamma_2) - \ln(c_1(t)) - \gamma_3\,\ln(c_2(t))\right],\\
      q(\ve c(t), \underbrace{p_1, p_2, p_3}_{\ve p}) &= p_1 - p_2\ln(c_1(t)) - p_3\ln(c_2(t)).
\end{align}
Here the parameters $\gamma_1$, $\gamma_2$, and $\gamma_3$ can be related to phenomenological constants; $\gamma_1$ stands for the mass-transfer coefficient, $\gamma_2$ is the limiting concentration of the macro-solute, and $\gamma_3$ is a dimensionless non-ideality factor. This model is proposed in~\cite{raj91} as a generalization of a limiting-flux model, which originates from film-theory of mass transfer~\citep{fick1855} and where $\gamma_3=0$. We study situations where the permeate flux at the plant obeys either one of these models.

As Eq.~\eqref{eq:ext_lim_model} suggests, an equivalent re-parameterization of the model is possible, which gives the model linear in parameters $p_1$, $p_2$, and $p_3$. This is convenient for parameter estimation. In this work, we will assume that the concentrations $c_1(t)$ and $c_2(t)$ can be measured perfectly (i.e., their measurement sensors are noise-free). As the values of state variables are known exactly, the dynamic equations can be eliminated from the problem~\eqref{eq:nlp_par_bounds}. Hence, the problem of estimating bounds of the model parameters~\eqref{eq:nlp_par_bounds} boils down to a problem of linear programming, since dynamics can be excluded and since the re-parameterization of the model yields linear-in-parameters structure of the model. The measurement noise associated with $q$ is assumed to be bounded $\sigma = 0.1\unit{L/h}$ and the measurements are available each second. In the simulation studies below, the realization of noise will be taken from a uniform distribution $\mathcal U(-\sigma, \sigma)$.

The objective is to find $u(t)$, which guarantees the transition from the given initial ($c_{1,0}$ and $c_{2,0}$) to final ($c_{1,\ff}$ and $c_{2,\ff}$) concentrations in minimum time. This problem can be formulated as:
\begin{subequations}\label{eq:t_opt_prob}
\begin{align}\label{eq:obj_func}
&\min_{t_\ff, u(t)}  \int_{0}^{t_\ff}\!1\,\dt,\\
\text{s.t. } \quad \dot c_1(t)&=\frac{c_1^2(t)q(\ve c(t), \ve p)}{c_{1,0}V_0}(1-u(t)), \quad
c_1(0)=c_{1,0},\quad c_1(t_{\ff}) =c_{1,\ff}, \label{eq:c1}\\
\dot c_2(t)&= - \frac{c_1(t) c_2(t)q(\ve c(t), \ve p)}{c_{1,0}V_0}u(t), \quad
c_2(0)=c_{2,0}, \quad c_2(t_{\ff}) =c_{2,\ff}, \label{eq:c2}\\
q(\ve c(t), \ve p) &= p_1 - p_2\ln(c_1(t)) - p_3\ln(c_2(t)),\\
u(t)&\in [0,\infty). \label{eq:uminmax}
\end{align}
\end{subequations}
The parameters of the problem are $c_{1,0} = 50\,\unit{g/L}$, $c_{1,\ff} = 150\,\unit{g/L}$, $c_{2,0} = 50\,\unit{g/L}$, $c_{2,\ff} = 0.05\,\unit{g/L}$, $V_0  = 20\,\unit{L}$, and $A=1\,\unit{m^2}$. Note that the extremal values of $u(t)$ in~\eqref{eq:uminmax} stand for a mode with no water addition, when $u(t)=0$ and pure dilution, \ie a certain amount of water is added at a single time instant, $u(t)=\infty$.

The parameterized optimal control of this process can be identified using Pontryagin's minimum principle~\citep{pon62} as~\eqref{eq:singular_ctrl_switch} where the singular control and the respective switching function can be found explicitly~\citep{pau12jms} as
\begin{align}
 u_s(\ve c(t,\ve p), \ve p) &:= \frac{1}{1+\gamma_3}=\frac{p_2}{p_2+p_3}, \\
 S(\ve c(t, \ve p), \ve p) &:= A\gamma_1\left(\ln(\gamma_2) - \ln(c_1)
 - \gamma_3\,\ln(c_2) - \gamma_3 - 1\right),\notag\\
 &:= \underbrace{p_1 - p_2\ln(c_1) - p_3\,\ln(c_2)}_{q(\ve c(t), \ve p)} - p_2 - p_3.
 \label{eq:sing_arc}
\end{align}
The structure of the optimal-control policy clearly reveals that the singular arc condition gives a constant value for the permeate flux (equal to $p_2+p_3$) and that the singular control is a constant that depends on the value of $\gamma_3=p_3/p_2$. This shows that if one devises a feedback-based real-time optimization scheme, precise estimation of parameters $p_2$ and $p_3$ is of paramount interest.

For the simulation-based studies on the implementation of the outlined optimal-control policy, we will assume that the nominal values of the parameters are $\gamma_1 = 3\times10^{-2}\,\unit{L/h}$, $\gamma_2 = 1000\,\unit{g/L}$, and $\gamma_3 = 0.1$. Similar values of the parameters were observed to validate the model against experimental data in~\cite{sha19}. The uncertainty at the initial point in time ($\ve P_0$) will be assumed as $\pm10\%$ of the nominal values. The true realization of the parameter values will be taken randomly from a uniform distribution $\mathcal U(p_0^L, p_0^U)$. As we deal with a time-minimization problem and the sampling time of the plant is 1 second, we naturally select the $\varepsilon=1\,\unit{s^2}$. For this simple example, the reachability analysis can be performed explicitly using the expressions for switching times provided in~\cite{fikar_2016}.

\subsection{Plant under limiting-flux conditions}
We first study the case when the plant is under limiting-flux conditions, i.e., the flux obeys Eq.~\eqref{eq:ext_lim_model} with $\gamma_3=p_3=0$. Based on the values of initial conditions and range of uncertainty in the parameters, the optimal-control policy boils down to three arcs:
\begin{enumerate}
 \item Use $u(t)=0$ until $t_1$, when $c_1(t_1)=\gamma_2/\text e$.
 \item Use $u(t)=u_s(t)=1$ until $c_1(t_2)/c_2(t_2)=c_{1,\ff}/c_{2,\ff}$.
 \item Dilute the solution (use $u(t)=\infty$ instantaneously) to arrive at the final concentrations.
\end{enumerate}
Taking into account that we measure both concentrations precisely, the only uncertainty in this case lies in the switching times $t_1$, where $t_1$ depends on the value of $\gamma_2$.

The implementation of the scheme, where one is aware of the true value of $\gamma_2$, results in $t_1^\text{opt}=2.515\,\unit h$ and $t_\ff^\text{opt}=8.284\,\unit h$. The worst-case minimization of the batch variability (methodology described in Section~\ref{sec:rob}) coincides in this case with the nominal strategy, where one takes $\gamma_2=\gamma_2^\text{nom}$. When applied to the plant, this strategy results in $t_1^\text{rob}=t_1^\text{nom}=2.625\,\unit h$ and $t_\ff^\text{rob}=t_\ff^\text{nom}=8.327\,\unit h$. The adaptive real-time dynamic optimization (described in Section~\ref{sec:adapt}) results in $t_1^\text{adapt}=2.533\,\unit h$ and $t_\ff^\text{adapt}=8.301\,\unit h$, which is only a slight improvement compared to the robust (and nominal) strategy.

\begin{figure}
\centering
\includegraphics[width=0.8\textwidth]{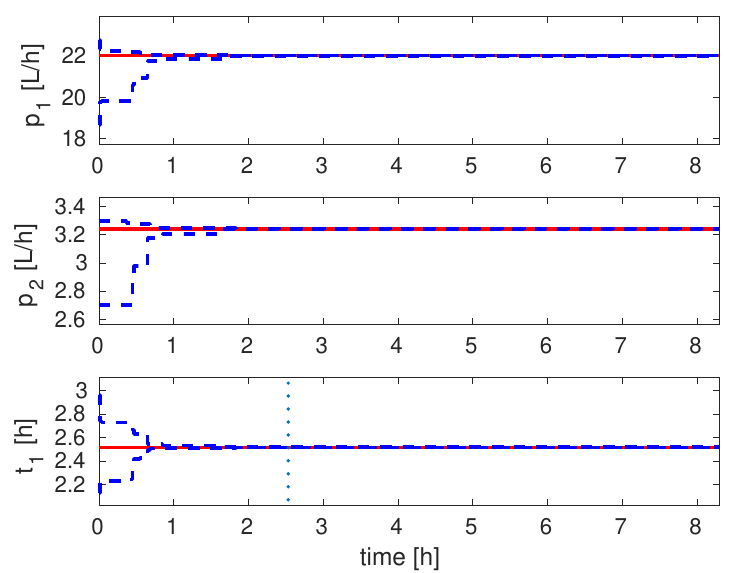}
\caption{Results of the set-membership estimation over time (top and middle plots) with projection of the uncertainty in the parameters on the switching time $t_1$ (bottom plot). The true (optimal) values are shown as solid lines, the bounds are represented using dashed lines. The vertical line in the bottom plot indicates the optimal switching time.}
\label{fig:perf_est_2par}
\end{figure}

Figure~\ref{fig:perf_est_2par} presents performance of the estimation (in terms of estimated parameter bounds) throughout the run of the batch. It is clear that the bounds on both parameters are dramatically reduced around the time point of 2\,h, which precedes the time point $t_1^\text{opt}$, when the switch in the control input should be executed. The bottom plot of Fig.~\ref{fig:perf_est_2par} also shows the evolution of the uncertainty in $t_1$, which is projected using interval-based calculations (as discussed in Section~\ref{ssec:param}). It should be noted here that the adaptive approach is successful mainly since the applied control input in the first arc coincides with an input that would result from a dynamic optimal-experiment design study. Here, $u(t)=0$ ensures the fastest possible increase of concentration $c_1(t)$, which reveals the most informative measurements about $\gamma_2$.

\begin{figure}
\centering
\includegraphics[width=0.7\textwidth]{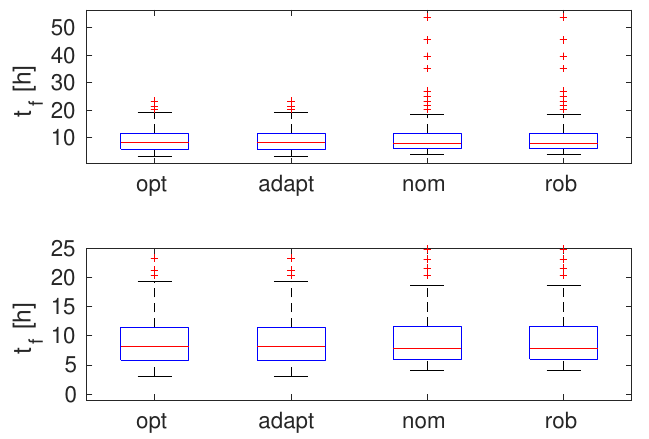}
\caption{A box plot with the statistical information (the median, the 25$^\text{th}$ and 75$^\text{th}$ percentiles and the outliers) about the performance of the different control strategies on the plant under limiting-flux conditions. The bottom plot shows a zoom of the top plot.}
\label{fig:perf_comp_2par}
\end{figure}

Finally, we evaluate a statistical performance of the presented dynamic real-time optimization schemes. This is realized by running 1,000 simulated batches with different true values of parameters. The resulting statistics is shown in Fig.~\ref{fig:perf_comp_2par}. One can clearly notice here that the robust and nominal strategies perform very well on average and even the standard deviations of their performance are not significantly increased compared to the optimal performance. On the other hand, distributions of the final batch times of the nominal and robust strategies have long tails, which points towards the existence of rare cases where the batch time obtained by application of robust dynamic real-time optimization increases significantly compared to the truly optimal solution. This behavior corresponds to the situations, where the first step (with $u(t)=0$) over-concentrates the solution too much so that the subsequent diafiltration step (with $u(t)=1$) requires long time to reach the desired condition ($c_1(t_2)/c_2(t_2)=c_{1,\ff}/c_{2,\ff}$).
This study reveals that the strategy, which uses estimation of parameter bounds, has a clear merit as it does not show this type of inconsistency in the performance (long tails) and it results overall in the batch times very close to the optimal ones.

Drawing a comparison in the computational time, robust and nominal strategies only require a single optimization before the batch starts. Adaptive strategy uses only a single re-optimization (scheduled just before $t_1^\text{nom}$) in this case, which shows a significant reduction in computational burden \wrt receding-horizon strategies. Due to explicit nature of the optimal control strategy, the computation of this re-optimization mostly lies in the estimation step. As the estimation problem can be boiled down to an LP, its solution is available in order of milliseconds using MATLAB's \emph{linprog} routine. Due to only a single re-optimization, one can also interpret this scheme as a mid-course correction~\citep{yab02} with optimally timed adaptation.

\subsection{Plant under generalized limiting-flux conditions}
Based on the values of initial conditions and range of uncertainty in the parameters, the optimal-control policy again boils down to three arcs:
\begin{enumerate}
 \item Use $u(t)=0$ until $t_1$, when~\eqref{eq:sing_arc} is zero.
 \item Use $u(t)=u_s(t)=1/(1+\gamma_3)$ until $c_1(t_2)/c_2(t_2)=c_{1,\ff}/c_{2,\ff}$.
 \item Dilute the solution (use $u(t)=\infty$ instantaneously) to arrive at the final concentrations.
\end{enumerate}
The uncertainty is in this case extended even on the value of singular control input and it is clear that one needs good estimates of both the values of $\gamma_2$ (which mostly influences the switching time $t_1$ as in the previous case) and $\gamma_3$ (which influences the quality of the singular control) to achieve a good performance. Knowledge of the a precise value of the parameter $\gamma_1$ is of minor importance as this parameter can be factored out of the optimality conditions.

We use the same values of the uncertain parameters $\gamma_1$ and $\gamma_2$ as in the previous case. The performance of the studied schemes is as follows:
\begin{itemize}
 \item Optimal strategy: $t_1^\text{opt}=2.501\,\unit h$, $t_\ff^\text{opt}=9.254\,\unit h$
 \item Adaptive strategy: $t_1^\text{adapt}=2.510\,\unit h$ and $t_\ff^\text{adapt}=9.271\,\unit h$
 \item Nominal strategy: $t_1^\text{nom}=2.561\,\unit h$ and $t_\ff^\text{nom}=9.277\,\unit h$
 \item Robust strategy: $t_1^\text{rob}=2.417\,\unit h$ and $t_\ff^\text{rob}=9.269\,\unit h$
\end{itemize}
We can observe a similar differences between the strategies as in the previous case. Robust strategy performs on an acceptable level and even marginally outperforms the nominal and the adaptive strategy, which results from the fact that the plant parameters coincide with the worst-case parameters.

\begin{figure}
\centering
\includegraphics[width=0.8\textwidth]{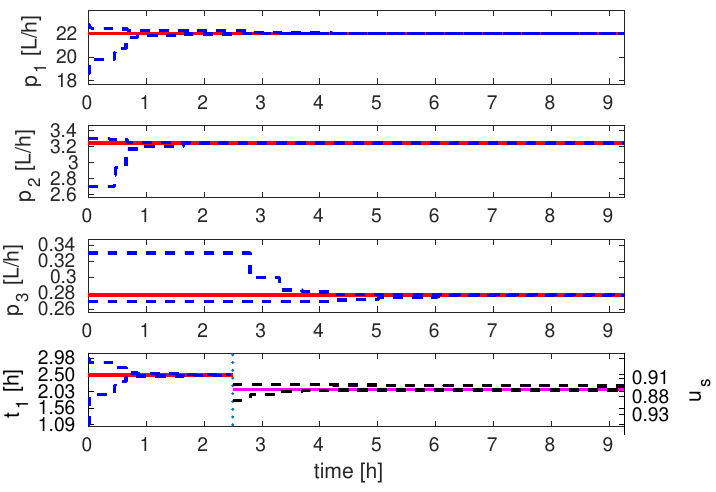}
\caption{Results of the set-membership estimation over time (top three plots) with projection of the uncertainty in the parameters on the switching time $t_1$ and on the value of $u_s$ (bottom plot). The true (optimal) values are shown as solid lines, the bounds are represented using dashed lines. The vertical line in the bottom plot indicates the optimal switching time.}
\label{fig:perf_est}
\end{figure}

Figure~\ref{fig:perf_est} presents performance of the estimation (in terms of estimated parameter bounds) throughout the run of the batch. Similarly to the previous case, the estimation performed in the first control arc helps in determination of the value of the first switching time before the optimal switching instant occurs. Here the determining parameter is $p_2$, whose estimation performance was discussed in the previous case and the same conclusions hold here. 

When the controller applies $u(t)=0$, the parameter $\gamma_3$ (or $p_3$) is unidentifiable as the concentration $c_2(t)$ remains constant. This can be seen in Fig.~\ref{fig:perf_est} as the bounds on $p_3$ remain constant from the beginning of the operation until the time when control input is switched to singular. It is also shown that the uncertainty in $p_3$ results in a relatively small uncertainty in the value of the singular control, so a precise knowledge of $p_3$ is not paramount for the application of the optimal control policy.

\begin{figure}
\centering
\includegraphics[width=0.7\textwidth]{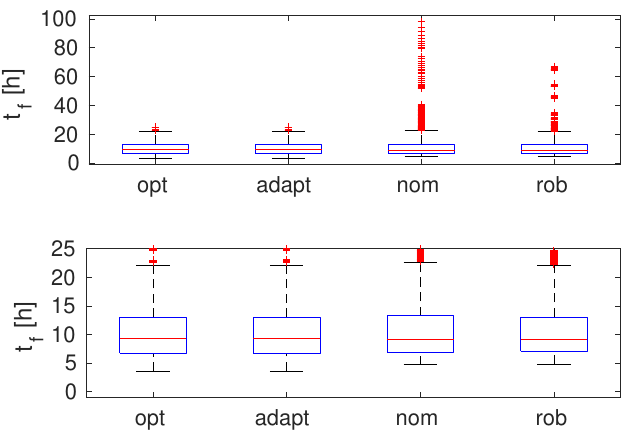}
\caption{A box plot with the statistical information (the median, the 25$^\text{th}$ and 75$^\text{th}$ percentiles and the outliers) about the performance of the different control strategies under generalized limiting-flux conditions. The bottom plot shows a zoom of the top plot.}
\label{fig:perf_comp}
\end{figure}

When the statistical performance is evaluated, we can first conclude that all the strategies perform almost identically on average. The biggest differences arise when one evaluates the outliers of the distribution of the achieved batch times. It is evident that the nominal strategy achieves the worst performance and that the robust strategy reduces the batch variability to a good extent (given the wide range of the uncertainty). The adaptive strategy is clearly superior here as it reduces the batch variability much further, compared to the robust scheme, and the achieved performance is practically indistinguishable from the truly optimal one.

\section{Conclusion}\label{sec:conclusions}
We have presented a methodology for dynamic real-time optimization of batch processes via parameterization of the optimal controller using Pontryagin's minimum principle. The employed parameterization greatly reduces the computational burden to guarantee feasibility of the operation compared to receding-horizon strategies. In order to address parametric plant-model mismatch issue, we have suggested a robust approach, which consisted in projection of the plant uncertainty into optimality conditions using reachability analysis. This again greatly reduces on-line computational burden as one can exploit the uncertainty in the switching to schedule the on-line re-optimization. As the uncertainty in parameters can greatly affect the optimality of the batch, we have proposed an adaptive scheme that makes use of parameter estimation and, as shown in the case study, can greatly assist in reducing variability in the batch performance subject to parametric uncertainty. The adaptive scheme turned out to be key in reduction of batch-to-batch variability. The future work will consider implementation of the proposed strategy on a laboratory plant.

\section*{Acknowledgments}
The authors gratefully acknowledge the contribution of the Scientific Grant Agency of the Slovak Republic under the grant 1/0004/17, of the Slovak Research and Development Agency under the project APVV 15-0007 and of the European Commission under the grant 790017 (GuEst). This publication is also a partial result of the Research \& Development Operational Programme for the project University Scientific Park STU in Bratislava, ITMS 26240220084, supported by the Research 7 Development Operational Programme funded by the ERDF. This work was also supported by the funding from Slovak Ministry of Education, Science, Research and Sport under the project STU as the Leader of Digital Coalition 002STU-2-1/2018.

\section*{Notation} \label{app:notation}
\begin{longtable}{cl}
$t$ & time [h]\\
$\ve x$ & vector of state variables\\
$\ve p$ & vector of model parameters\\
$u$ & manipulated variable\\
$\mathcal J$ & objective functional\\
$F_0$ & constant-in-control part of the Lagrange term\\
$F_u$ & multiplier of linear-in-control part of the Lagrange term\\
$\ve f_0$ & constant-in-control term of the model\\
$\ve f_u$ & multiplier of linear-in-control term of the model\\
$\ve P$ & interval box\\
$H$ & Hamiltonian (function)\\
$H_0$ & constant-in-control term of the Hamiltonian\\
$H_u$ & multiplier of linear-in-control term of the Hamiltonian\\
$\mu$ & Lagrange multiplier of bound on manipulated variable\\
$\ve\nu$ & Lagrange multiplier of final conditions\\
$\ve\lambda(\cdot)$ & vector of adjoint variables\\
$S$ & state-dependent switching function\\
$\ve\pi$ & vector parameterizing the optimal control strategy\\
$\ve g$ & measurement function\\
$\ve y$ & vector of predicted plant outputs\\
$\ve y_p(t)$ & vector of true outputs of the plant\\
$\ve y_m(t)$ & vector of measured outputs\\
$\ve\sigma$ & magnitude of measurement noise\\
mid & mid-point of the interval box\\
$\varepsilon$ & user-defined tolerance\\
$A$ & membrane area [m$^2$]\\
$V$ & volume of the processed solution\\
$c_1$ & macro-solute concentration\\
$c_2$ & micro-solute concentration\\
$q$ & permeate flux, flow rate through the membrane\\
$\gamma_1$ & mass-transfer coefficient\\
$\gamma_2$ & limiting concentration of the macro-solute\\
$\gamma_3$ & dimensionless non-ideality factor\\
$\ve\gamma$ & parameters of the original permeate-flux model\\
$p_1$, $p_2$, $p_3$ & parameters of the reparametrized permeate-flux model\\
$\mathcal U(a, b)$ & uniform distribution bounded by $a$ and $b$\\

\multicolumn{2}{l}{\itshape Subscripts}\\
0 & initial, control-independent\\
$u$ & linear-in-control part\\
$k$ & sampling instant of the plant\\
$\ff$ & final\\
$s$ & singular\\

\multicolumn{2}{l}{\itshape Superscripts}\\
$L$ & lower bound\\
$U$ & upper bound\\
opt & optimal\\
nom & nominal\\
adapt & adaptive\\
rob & robust
\end{longtable}

\bibliographystyle{elsarticle-harv}\biboptions{authoryear}

\end{document}